\documentclass[12pt]{article}
\usepackage[left=2.5cm,top=2.50cm,right=2.5cm,bottom=2.50cm]{geometry}
\usepackage{mathrsfs}
\usepackage{amsmath,amssymb,latexsym,color,cancel,graphicx,colortbl}
\usepackage[english]{babel}
\usepackage[latin1]{inputenc}
\usepackage{cite}
\begin{document}
\date{}

\title{Algebraic approach to the Tavis-Cummings model with three modes of oscillation}
\author{E. Chore$\tilde{n}$o$^{a}$, \footnote{{\it E-mail address:} echorenoo0800@alumno.ipn.mx}\\ D. Ojeda-Guill\'en$^{b}$ and V. D. Granados$^{a}$} \maketitle

\begin{minipage}{0.9\textwidth}
\small $^{a}$ Escuela Superior de F{\'i}sica y Matem\'aticas,
Instituto Polit\'ecnico Nacional, Ed. 9, Unidad Profesional Adolfo L\'opez Mateos, Delegaci\'on Gustavo A. Madero, C.P. 07738, Ciudad de M\'exico, Mexico.\\
\small $^{b}$ Escuela Superior de C\'omputo, Instituto Polit\'ecnico Nacional,
Av. Juan de Dios B\'atiz esq. Av. Miguel Oth\'on de Mendiz\'abal, Col. Lindavista,
Delegaci\'on Gustavo A. Madero, C.P. 07738, Ciudad de M\'exico, Mexico.\\

\end{minipage}

\begin{abstract}

We study the Tavis-Cummings model with three modes of oscillation by using four different algebraic methods: the Bogoliubov transformation, the normal-mode operators, and the tilting transformation of the $SU(1,1)$ and $SU(2)$ groups. The algebraic method based on the Bogoliubov transformation and the normal-mode operators let us obtain the energy spectrum and eigenfunctions of a particular case of the Tavis-Cummings model, while with the tilting transformation we are able to solve the most general case of this Hamiltonian. Finally, we compute some expectation values of this problem by means of the $SU(1,1)$ and $SU(2)$ group theory.

\end{abstract}

\section{Introduction}

The Jaynes-Cummings model was introduced to describe the interaction between a two-level atom with a quantized electromagnetic field \cite{Jay}.
This is the simplest completely soluble quantum-mechanical model and is an intrinsically nonlinear model. In the rotating wave approximation this model have been extensively studied and its exact solution has been found \cite{Haroche}. These solutions yield quantum collapse and revival of atomic inversion \cite{Narozhny}, and squeezing of the radiation field \cite{Kuklinski}, among other quantum effects. All these effects have been corroborated experimentally, as can be seen in references \cite{Goy,Brune,Guerlin}. Another important fact to highlight of the model is that it has been possible to perform experimentally \cite{Meschede,Rempe,Rempe2,An,Schmidt}. In references \cite{NosJMP,NosAP}, generalizations of the Jaynes-Cummings model have been introduced with one and two modes of oscillation in terms of the creation and annihilation operators of the one-dimensional harmonic oscillator. As physical applications, these generalizations were connected with the relativistic parametric amplifier, the quantum simulation of a single trapped ion and the relativistic problem of two coupled oscillators.

On the other hand, Dicke pointed out the importance of treating a gas of radiating molecules as a single quantum system, where the molecules are
interacting with a common radiation field \cite{Dicke}. Tavis and Cummings studied and exactly solved the problem of $N$ identical two-level molecules interacting through a dipole coupling with a single-mode quantized radiation field at resonance \cite{TC}. In their solution, Tavis and Cummings used the so-called rotating-wave approximation to ignore some terms in the dipole coupling which do not conserve energy in the first order of perturbation. The Tavis-Cummings model is a generalization of the Jaynes-Cummings model and, due to historical reasons, the Tavis-Cummings model is often called the Dicke model.

To study the Tavis-Cummings model and some of its modifications various algebraic methods have been used, as the Holstein-Primakoff transformation \cite{Bashir}, quantum inverse methods \cite{Bogoliubov,Rybin}, and polynomially deformed $su(2)$ algebras \cite{Vadeiko}. However, it has been showed that under certain considerations, the rotating-wave approximation fails in the description of the phenomena and cannot be neglected \cite{Milonni,Agarwal}.
The Jaynes-Cummings model and its generalization given by the Tavis-Cummings model are still under study as can be seen in references \cite{Lamata,Gerritsma,Lamata2,Retzker,Kopylov,Sun}. In particular, these two models have been used in quantum information science in the study of circuit-QED \cite{Romero,Schotland,Fink}.

The aim of the present work is to exactly solve the Tavis-Cummings model with three modes of oscillation by using four different algebraic methods: the Bogoliubov transformation, the normal-mode operators, and the tilting transformation of the $SU(1,1)$ and $SU(2)$ groups.

This work is organized as follows. In Section 2, we study the Tavis-Cummings model with three modes of oscillation by using the Bogoliubov transformation and the $SU(1,1)$ tilting transformation. The first algebraic method let us obtain the energy spectrum of a particular case of the Tavis-Cummings mode, while the tilting transformation let us obtain the energy spectrum and eigenfunctions of the more general model. In Section 3, we proceed as in the previous Section and study the Tavis-Cummings model by introducing the normal-mode operators and the $SU(2)$ tilting transformation. In Section 4, we compute the expectation value of the Tavis-Cummings Hamiltonian in terms of the coherent states of the one-dimensional harmonic oscillator, and the coherent states of the $SU(1,1)$ and the $SU(2)$ groups. Then, we make a comparison between these results. Finally, we give some concluding remarks.

\section{$SU(1,1)$ approach to the Tavis-Cummings model}

The interaction between three electromagnetic fields represents an important nonlinear parametric interaction. This interaction is of nonlinear type and can be transformed into either parametric amplifier or parametric frequency converter; the first type leads to amplification of the system energy while the second type leads to the energy exchanges between modes. The most familiar Hamiltonian representing such a system is given by \cite{Abdalla}
\begin{equation}
H = \omega_{1}\hat{a}^{\dag}\hat{a}+\omega_{2}\hat{b}^{\dag}\hat{b}+\omega_{3}\hat{c}^{\dag}\hat{c}+g(\hat{a}^{\dag}\hat{b}\hat{c}+\hat{a}\hat{b}^{\dag}\hat{c}^{\dag}),\label{T-C}
\end{equation}
where we set $\hbar=1$, $\omega_{j}$ with $j=1,2,3$ are the field frequencies, and $g$ is the coupling constant. The set of operators $\hat{a},\hat{a}^{\dag}$, $\hat{b},\hat{b}^{\dag}$ and $\hat{c},\hat{c}^{\dag}$ satisfy the bosonic algebra
\begin{equation}
[a,a^{\dag}]=[b,b^{\dag}]=[c,c^{\dag}]=1.
\end{equation}
The Hamiltonian of equation (\ref{T-C}) can be seen as the Tavis-Cummings model with three modes of oscillation. In this Section, we shall give an algebraic solution of a particular case of this model by using the Bogoliubov transformation. The most general Hamiltonian will be treated by means of the tilting transformation of the $su(1,1)$ Lie algebra.

\subsection{The Tavis-Cummings model and the Bogoliubov transformation}

A particular case of the Tavis-Cummings model is obtained if we set $\omega_{2}=\omega_{3}=\omega$. Moreover, if we define $\hat{\beta}=g\hat{a}$ and $\hat{\beta}^{\dag}=g^{*}\hat{a}^{\dag}$, the Hamiltonian of equation (\ref{T-C}) can be written as
\begin{equation}
H = \omega_{1}\hat{a}^{\dag}\hat{a}+\omega(\hat{b}^{\dag}\hat{b}+\hat{c}^{\dag}\hat{c})+\hat{\beta}^{\dag}\hat{b}\hat{c}+\hat{\beta}\hat{b}^{\dag}\hat{c}^{\dag}.\label{TC2}
\end{equation}

Now, we introduce the Bogoliubov transformation for two modes
\begin{equation*}
\hat{b}=\hat{\bar{a}}\cosh{r}+\hat{d}^{\dag}e^{-i\theta}\sinh{r},
\end{equation*}
\begin{equation}
\hat{c}=\hat{d}\cosh{r}+\hat{\bar{a}}^{\dag}e^{-i\theta}\sinh{r},
\end{equation}
where the operators $\hat{\bar{a}},\hat{\bar{a}}^{\dag}$ and $\hat{d},\hat{d}^{\dag}$ also satisfy the bosonic algebra. Thus, in terms of these operators the Hamiltonian of the equation (\ref{TC2}) becomes
\begin{align}
H'=& \left[\omega(1+2\sinh^{2}r)+\frac{1}{2}(\hat{\beta}e^{i\theta}+\hat{\beta}^{\dag}e^{-i\theta})\sinh{2r}\right]\left(\hat{\bar{a}}^{\dag}\bar{a}+\hat{d}^{\dag}\hat{d}+1\right)\nonumber\\&+ \left[e^{i\theta}\sinh{2r}+\hat{\beta}^{\dag}\cosh^{2}{r}+\hat{\beta}e^{2i\theta}\sinh^{2}r\right]\hat{\bar{a}}\hat{d}\nonumber\\&+ \left[e^{-i\theta}\sinh{2r}+\hat{\beta}^{\dag}e^{-2i\theta}\sinh^{2}{r}+\hat{\beta}\cosh^{2}r\right]\hat{\bar{a}}^{\dag}\hat{d}^{\dag}\nonumber\\&+\omega_{1}\hat{a}^{\dag}\hat{a}-\omega.\label{TBogo}
\end{align}
The operators $\hat{\bar{a}}\hat{d}$ and $\hat{\bar{a}}^{\dag}\hat{d}^{\dag}$ can be removed if the parameters $r,\theta$ satisfy the following equations
\begin{equation}
e^{i\theta}\sinh{2r}+\hat{\beta}^{\dag}\cosh^{2}{r}+\hat{\beta}e^{2i\theta}\sinh^{2}r=0,
\end{equation}
\begin{equation}
e^{-i\theta}\sinh{2r}+\hat{\beta}^{\dag}e^{-2i\theta}\sinh^{2}{r}+\hat{\beta}\cosh^{2}r=0.
\end{equation}
From these equations we find that the parameters $r,\theta$ explicitly are
\begin{equation}
\hat{r}=\frac{1}{2}\ln\left[\frac{\omega+\sqrt{\hat{\beta}^{\dag}\hat{\beta}}}{\omega-\sqrt{\hat{\beta}^{\dag}\hat{\beta}}}\right],  \quad\quad \hat{\theta}=i\ln\left[\sqrt{\frac{\hat{\beta}}{\hat{\beta}^{\dag}}}\right].\label{parameters}
\end{equation}
Note that the operator $\frac{1}{\hat{\beta}^{\dag}}$ is the inverse operator of $\hat{\beta}^{\dag}$, i.e.
\begin{equation}
\frac{1}{\hat{\beta}^{\dag}}\hat{\beta}^{\dag}=1,  \quad\quad \hat{\beta}^{\dag}\frac{1}{\hat{\beta}^{\dag}}=1-|0\rangle\langle{0}|.
 \end{equation}
By substituting the parameters of equations (\ref{parameters}) into equation (\ref{TBogo}), we obtain that the Hamiltonian of the Tavis-Cummings model is diagonalized as
\begin{equation}
H'=\omega_{1}\hat{a}^{\dag}\hat{a}+\sqrt{\omega^{2}-g\hat{a}^{\dag}\hat{a}}\left(\hat{\bar{a}}^{\dag}\bar{a}+\hat{d}^{\dag}\hat{d}+1\right)-\omega.\label{TCBolo}
\end{equation}
Therefore, the energy spectrum of the Tavis-Cummings model of equation (\ref{TC2}) from the Bogoliubov transformation is
\begin{equation}
E_{BT}=\sqrt{\omega^2-g^2n_a}\left(n_{\bar{a}}+n_d+1\right)+\omega_1n_a-\omega.
\end{equation}

Let us now look the eigenfunctions $\Psi'$ of this Tavis-Cummings model. Since the operator $\hat{\bar{a}}^{\dag}\bar{a}+\hat{d}^{\dag}\hat{d}+1$ is the Hamiltonian of the two-dimensional harmonic oscillator and commutes with $\hat{a}^{\dag}\hat{a}$, we have that the eigenfunctions of this Hamiltonian are given as the direct product
\begin{equation}
\Psi'=\psi_{n_a}(x)\otimes\psi_{n_{l},m_n}(\rho,\phi).
\end{equation}
In this expression $\psi_{n_a}(x)$ are the eigenfunctions of the one-dimensional harmonic oscillator
\begin{equation}
\psi_{n_a}(x)=\langle{x}|n_{a}\rangle=\sqrt{\frac{1}{\pi^{1/4}(2^n n!)^{1/2}}}e^{-\frac{1}{2}x^2}H_n(x),
\end{equation}
and $\psi_{n_{l},m_n}(\rho,\phi)$ are the eigenfunctions of the two-dimensional harmonic oscillator
\begin{equation}
\psi_{n_{l},m_n}(\rho,\phi)=\langle\rho,\phi|n_{l},m_{n}\rangle=\frac{1}{\sqrt{\pi}}e^{im_n\phi}(-1)^{n_{l}}\sqrt{\frac{2(n_{l})!}{(n_{l}+m_n)!}}\rho^{m_n}L^{m_n}_{n_{l}}(\rho^{2})e^{-1/2\rho^{2}},\label{Poly1}
\end{equation}
with the left chiral quantum number $n_{l}$=$(n-m_n)/2$. Here, $H_n(x)$ are the Hermite polynomials and $L^{m_n}_{n_{l}}(\rho^{2})$ are the Laguerre polynomials.

Therefore, we have solved a particular case of the Tavis-Cummings model (with the field frequencies $\omega_{2}=\omega_{3}=\omega$) by using the Bogoliubov transformation for two modes. We obtained the energy spectrum and its transformed eigenfunctions in terms of the one and two-dimensional harmonic oscillator.

\subsection{The Tavis-Cummings model and the $SU(1,1)$ tilting transformation}

In order to use the $SU(1,1)$ group theory, we can rewritte the general Hamiltonian of the Tavis-Cummings model of equation (\ref{T-C}) as
\begin{equation}
H=\omega_{1}\hat{a}^{\dag}\hat{a}+\frac{(\omega_{2}+\omega_{3})}{2}(\hat{b}^{\dag}\hat{b}+\hat{c}^{\dag}\hat{c})+\frac{(\omega_{2}-\omega_{3})}{2}(\hat{b}^{\dag}\hat{b }-\hat{c}^{\dag}\hat{c})+g(\hat{a}^{\dag}\hat{b}\hat{c}+\hat{a}\hat{b}^{\dag}\hat{c}^{\dag}).
\end{equation}
Now, we introduce the Jordan-Schwinger realization of the $su(1,1)$ algebra in terms of the operators $\hat{b}$ and $\hat{c}$ (see equations (\ref{gen11}) and (\ref{Nd}) of Appendix)
\begin{equation}\nonumber
K_0=\frac{1}{2}\left(\hat{b}^{\dag}\hat{b}+\hat{c}^{\dag}\hat{c}+1\right), \quad\quad K_+=\hat{b}^{\dag}\hat{c}^{\dag},
\end{equation}
\begin{equation}
K_-= \hat{b}\hat{c},\quad\quad N_d=\hat{c}^{\dag}\hat{c}-\hat{b}^{\dag}\hat{b}.
\end{equation}
Thus, the eigenvalue equation of the Tavis-Cummings model described by the Jordan-Schwinger realization of the $su(1,1)$ Lie algebra becomes
\begin{align}
H_{su(1,1)}|\Psi\rangle &=\left[\omega_{1}\hat{a}^{\dag}\hat{a}+(\omega_{2}+\omega_{3})K_{0}+g(\hat{a}^{\dag}K_{-}+\hat{a}K_{+})+\frac{(\omega_{2}-\omega_{3})}{2}N_{d}-\frac{(\omega_{2}+\omega_{3})}{2}\right]|\Psi\rangle \nonumber \\&=E_{su(1,1}|\Psi\rangle.\label{Hsu11}
\end{align}

In this case, we can remove the ladder operators $K_{\pm}$ in this Hamiltonian $H_{su(1,1)}$ by using the tilting transformation in terms of the $SU(1,1)$ displacement operator $D(\xi)$. Therefore, if we apply the tilting transformation to both sides of equation (\ref{Hsu11}) and we proceed as in references \cite{gerryberry,Nos1,Nos2}, we obtain
\begin{equation*}
D^{\dag}(\xi)H_{su(1,1)}D(\xi)D^{\dag}(\xi)|\Psi_{1}\rangle =E_{su(1,1)}D^{\dag}(\xi)|\Psi_{1}\rangle,
\end{equation*}
\begin{equation}
H'|\Psi_{1}'\rangle=E_{su(1,1)}|\Psi_{1}'\rangle.
\end{equation}
Notice that in these expressions $H'=D^{\dag}(\xi)H_{su(1,1)}D(\xi)$ is the tilted Hamiltonian and $|\Psi_{1}'\rangle=D^{\dag}(\xi)|\Psi_{1}\rangle$ its wave function. Thus, by using equations (\ref{DK+}), (\ref{DK-}) and (\ref{DK0}) of Appendix, the tilted Hamiltonian can be written as
\begin{align}
H'=&\left[(\omega_{2}+\omega_{3})(2\beta+1)+g\hat{a}\frac{\xi^{*}\alpha}{|\xi|}+g\hat{a}^{\dag}\frac{\xi\alpha}{|\xi|}\right]K_{0}\nonumber\\&+ \left[(\omega_{2}+\omega_{3})\frac{\alpha\xi}{2|\xi|}+g\hat{a}^{\dag}\frac{\beta\xi}{\xi^{*}}+ g\hat{a}(\beta+1)\right]K_{+}\nonumber\\& +\left[(\omega_{2}+\omega_{3})\frac{\alpha\xi^{*}}{2|\xi|}+g\hat{a}^{\dag}(\beta+1)+g\hat{a}\frac{\beta\xi^{*}}{\xi}\right]K_{-}\nonumber\\&+\omega_{1}\hat{a}^{\dag}\hat{a}+\frac{(\omega_{2}-\omega_{3})}{2}\hat{N}_{d}-\frac{(\omega_{2}+\omega_{3})}{2}.\label{H2}
\end{align}
In this expression, the terms $N_d=\hat{b}^{\dag}\hat{b}-\hat{c}^{\dag}\hat{c}$ and $\hat{a}^{\dag}\hat{a}$ remain unchanged under the tilting transformation, since both operators commute with $K_{\pm}$ (see Appendix). By choosing the coherent state parameters $\theta$ and $\varphi$ as
\begin{equation*}
\hat{\theta}=\tanh^{-1}\left(\frac{2g\sqrt{\hat{a}^{\dag}\hat{a}}}{\omega_{2}+\omega_{3}}\right),\quad\quad\hat{\varphi}=i\ln{\left[\frac{2(2\beta+1)g\hat{a}}{\alpha(\omega_{2}+\omega_{3})} \right]},
\end{equation*}
the tilted Hamiltonian $H'$ of equation (\ref{H2}) is reduced to
\begin{equation}
H'=\sqrt{(\omega_{2}+\omega_{3})^{2}-4g^{2}\hat{a}^{\dag}\hat{a}}\hat{K}_{0}+\omega_{1}\hat{a}^{\dag}\hat{a}+\frac{(\omega_{2}-\omega_{3})}{2}\hat{N}_{d}-\frac{(\omega_{2}+\omega_{3})}{2}.\label{TC-til}
\end{equation}

On the other hand, since the operator $K_{0}$ is the Hamiltonian of the two-dimensional harmonic oscillator and commutes with the operators $\hat{N}_d$ and $\hat{a}^{\dag}\hat{a}$, the eigenfunction of $H'$ is given by the direct product
\begin{equation}
\Psi'=\psi_{n_a}(x)\otimes\psi_{n_{l},m_n}(\rho,\phi),\label{Poly2}
\end{equation}
where
\begin{equation}
\psi_{n_a}(x)=\langle{x}|n_{a}\rangle=\sqrt{\frac{1}{\pi^{1/4}(2^n n!)^{1/2}}}e^{-\frac{1}{2}x^2}H_n(x),
\end{equation}
and
\begin{equation}
\psi_{n_{l},m_n}(\rho,\phi)=\langle\rho,\phi|n_{l},m_{n}\rangle=\frac{1}{\sqrt{\pi}}e^{im_n\phi}(-1)^{n_{l}}\sqrt{\frac{2(n_{l})!}{(n_{l}+m_n)!}}\rho^{m_n}L^{m_n}_{n_{l}}(\rho^{2})e^{-1/2\rho^{2}}.
\end{equation}

From the action of the operators $\hat{c}$, $\hat{c}^{\dag}$, $\hat{b}$ and $\hat{b}^{\dag}$ on the basis $|n,m_n\rangle$, we have
\begin{equation}
K_{0}|n,m_n\rangle=\left(n_l+\frac{m_n}{2}+\frac{1}{2}\right)|n,m_n\rangle,\nonumber
\end{equation}
\begin{equation}
N_d|n,m_n\rangle=(\hat{b}^{\dag}\hat{b}-\hat{c}^{\dag}\hat{c})|n,m_n\rangle=-m_n|n,m_n\rangle.
\end{equation}
In this $SU(1,1)$ representation, the group numbers $n,k$ are related with the physical numbers $n_l,m_n$ as $n=n_l$ and $k=\frac{1}{2}(m_n+1)$ \cite{Nos1}. Thus, from these results we obtain that the energy spectrum of the most general case of the Tavis-Cummings model is
\begin{equation}
E_{su(1,1)}=\sqrt{(\omega_{2}+\omega_{3})^{2}-4g^{2}n_{a}}\left(n_l+\frac{m_n}{2}+\frac{1}{2}\right)+\omega_{1}n_{a}+\frac{(\omega_{3}-\omega_{2})}{2}m_{n}-\frac{(\omega_{2}+\omega_{3})}{2}.\label{hkg3}
\end{equation}

Notice that the Hamiltonian of equation (\ref{TC-til}) can be easily reduced to the isotropic three-dimensional harmonic oscillator when $\omega_{1}=\omega_{2}=\omega_{3}$ and $g=0$. If we substitute this expressions in the energy spectrum of equation (\ref{hkg3}), we obtain
the correct energy spectrum. In the same manner, when $\omega_{2}=\omega_{3}$ this Hamiltonian is simplified to the Hamiltonian of equation (\ref{TCBolo}) with the corresponding energy spectrum
\begin{equation}
E=\sqrt{\omega^{2}-g^{2}n_{a}}(n+1)+\omega_{1}n_{a}-\omega.
\end{equation}

The eigenfunctions $\Psi$ for the Hamiltonian $H_{su(1,1)}$ of the Tavis-Cummings model are obtained from equation (\ref{Poly2}). Thus, $\Psi=D(\xi)\Psi'=\psi_{a}(x)\otimes{D(\xi)\psi'_{n_{l},m_n}}(\rho,\phi)$. By using equation (\ref{PNCS}) of Appendix, we find that the action of the displacement operator $D(\xi)$ on $\psi'_{n_{l},m_n}(\rho,\phi)$ are the Perelomov number coherent states for the two-dimensional harmonic oscillator $\psi_{\zeta,n_{l},k}$ \cite{Nos1}
\begin{align*}
\psi_{\zeta,n_{l},k}=\langle\rho,\phi|\zeta,k,n_{l}\rangle= &\frac{(-1)^{n_{l}}}{\sqrt{\pi}}e^{i(l-1/2)\phi}\sum_{s=0}^{\infty}\frac{\zeta^{s}}{s!}\sum_{j=0}^{n_{l}}\frac{(-\zeta^{*})^{j}}{j!}e^{\eta(k+n_{l}-j)}\frac{\sqrt{2\Gamma(n_{l}+1)\Gamma\left(n_{l}+l+\frac{1}{2}\right)}}{\Gamma\left(n_{l}-j+l+\frac{1}{2}\right)}\nonumber\\&\times\frac{\Gamma(n_{l}-j+s+1)}{\Gamma(n_{l}-j+1)}e^{-\rho^{2}/2}\rho^{l-1/2}L_{n_{l}-j+s}^{l-1/2}(\rho^{2}),\
\end{align*}
with $l=m_n+\frac{1}{2}$.
The above expression can also be rewritten as
\begin{align}
\psi_{\zeta,n_{l},m_n}=&\sqrt{\frac{2\Gamma(n_{l}+1)}{\Gamma(n_{l}+m_n+1)}}\frac{(-1)^{n_{l}}}{\sqrt{\pi}}e^{im_n\phi}\frac{(-\zeta^{*})^{n_{l}}(1-|\zeta|^{2})^{\frac{m_n}{2}+\frac{1}{2}}(1+\sigma)^{n_{l}}}{(1-\zeta)^{m_n+1}}\nonumber\\&\times e^{-\frac{\rho^{2}(\zeta+1)}{2(1-\zeta)}}\rho^{m_n}L_{n_{l}}^{m_m}\left(\frac{\rho^{2}\sigma}{(1-\zeta)(1-\sigma)}\right),\label{eigen2}
\end{align}
where we have used $m_n=l-\frac{1}{2}$ and defined $\sigma$ as
\begin{equation*}
\sigma=\frac{1-|\zeta|^{2}}{(1-\zeta)(-\zeta^{*})}.
\end{equation*}
In this way, the functions of the Tavis-Cummings model are given by the direct product of the functions of the one-dimensional harmonic oscillator $\psi_{n_a}(x)$ and the $SU(1,1)$ Perelomov number coherent states for the two-dimensional harmonic oscillator $\psi_{\zeta,n_{l},m_n}$ as follows
\begin{equation}
\Psi=\psi_{n_a}(x)\otimes\psi_{\zeta,n_{l},m_n}.
\end{equation}

Therefore, the Tavis-Cummings model was solved by using the tilting transformation and the $SU(1,1)$ Perelomov number coherent states. Moreover, the solution obtained from this method coincides with the solution obtained by using the Bogoliubov transformation when we set $\omega_{2}=\omega_{3}$.

\section{$SU(2)$ approach to the Tavis-Cummings model}

In this Section, we shall obtain the solution of the Tavis-Cummings model with three modes of oscillation by using two methods: the normal-mode operators and the $SU(2)$ tilting transformation. As in Section 2, the normal-mode operators method will allow us to solve a particular case of the Tavis-Cummings model, while with the $SU(2)$ tilting transformation method we obtain the most general solution.

\subsection{The Tavis-Cummings model and the normal-mode operators}

For convenience, in this Section we shall use the Hamiltonian of the Tavis-Cummings model described in the interaction picture. To do this, we fist note that the Hamiltonian of equation (\ref{T-C}) can be written as
\begin{equation}
H=\omega_{3}\hat{c}^{\dag}\hat{c}+(\omega_1-\omega_2)\hat{a}^{\dag}\hat{a}+\omega_2\hat{N}_{s}+g(\hat{c}\hat{a}^{\dag}\hat{b}+\hat{c}^{\dag}\hat{a}\hat{b}^{\dag})\label{iTcx},
\end{equation}
where $\hat{N}_{s}=\hat{b}^{\dag}\hat{b}+\hat{a}^{\dag}\hat{a}$ is the number operator. Therefore, since $\hat{N}_{s}$ commute with the rest of the operators, the Hamiltonian in the interaction picture is \cite{Dutra}
\begin{equation}
H^{(i)}_{NM}=\omega_{3}\hat{c}^{\dag}\hat{c}+\Delta\hat{a}^{\dag}\hat{a}+g(\hat{c}\hat{a}^{\dag}\hat{b}+\hat{c}^{\dag}\hat{a}\hat{b}^{\dag})\label{iTc3},
\end{equation}
with $\Delta=(\omega_{1}-\omega_{2})$, the detuning. With the normal-mode operators defined as
\begin{equation}
\hat{A}_{1}=X\hat{a}+Y^{*}\hat{b},  \quad\quad \hat{A}_{2}=Y\hat{a}-X^{*}\hat{b}, \label{N-M}
\end{equation}
where $X$ and $Y$ are two constants to be determined, we find that the products $\hat{A}_{1}^{\dag}\hat{A}_{1}$ and $\hat{A}_{2}^{\dag}\hat{A}_{2}$ result to be
\begin{equation}
\hat{A}_{1}^{\dag}\hat{A}_{1}=|X|^{2}\hat{a}^{\dag}\hat{a}+X^{*}Y^{*}\hat{a}^{\dag}\hat{b}+XY\hat{a}\hat{b}^{\dag}+|Y|^{2}\hat{b}^{\dag}\hat{b},
\end{equation}
\begin{equation}
\hat{A}_{2}^{\dag}\hat{A}_{2}=|Y|^{2}\hat{a}^{\dag}\hat{a}-X^{*}Y^{*}\hat{a}^{\dag}\hat{b}-XY\hat{a}\hat{b}^{\dag}+|X|^{2}\hat{b}^{\dag}\hat{b}.
\end{equation}
Thus, we can relate these normal-mode operators with the Hamiltonian $\hat{H}^{(i)}_{NM}$ of equation (\ref{iTc3}) as
\begin{equation}
H^{(i)}_{NM}=\omega_{3}\hat{c}^{\dag}\hat{c}+\alpha\hat{A}_{1}^{\dag}\hat{A}_{1}+\beta\hat{A}_{2}^{\dag}\hat{A}_{2},
\end{equation}
if the conditions
\begin{equation}
|X|^{2}\alpha+|Y|^{2}\beta=\Delta, \quad\quad |Y|^{2}\alpha+|X|^{2}\beta=0,\label{xy}
\end{equation}
and
\begin{equation}
Y^{*}X^{*}(\alpha-\beta)=g\hat{c}, \quad\quad YX(\alpha-\beta)=g\hat{c}^{\dag},
\end{equation}
are satisfied. From these equations it is easy to see that
\begin{equation}
\Omega^{2}|X|^{2}|Y|^{2}=g^{2}\hat{c}^{\dag}\hat{c}.
\end{equation}
Hence, from this relation along with equations (\ref{xy}), we find that the values of the squared norm of $X$ and $Y$ are
\begin{equation}
|X|^{2}=\frac{2g^{2}\hat{c}^{\dag}\hat{c}}{\Omega\left(\Omega-\Delta\right)}, \quad\quad |Y|^{2}=\frac{\Omega-\Delta}{2\Omega},
\end{equation}
where $\Omega=\alpha-\beta=\sqrt{\Delta^{2}+4g^{2}\hat{c}^{\dag}\hat{c}}$.

On the other hand, the operators $\hat{A}_{1},\hat{A}_{1}^{\dag}$ and $\hat{A}_{2},\hat{A}_{2}^{\dag}$ satisfy the bosonic algebra
\begin{equation}
[\hat{A}_{1},\hat{A}_{1}^{\dag}]=[\hat{A}_{2},\hat{A}_{2}^{\dag}]=1,
\end{equation}
and commute with each other
\begin{equation}
[\hat{A}_{1},\hat{A}_{2}]=[\hat{A}_{1}^{\dag},\hat{A}_{2}^{\dag}]=0.
\end{equation}
In this way, the normal-mode operators can be considered as creation and annihilation operators and the operators $\hat{A}_{1}^{\dag}\hat{A}_{1}$ and $\hat{A}_{2}^{\dag}\hat{A}_{2}$ as number operators. Thus, the Hamiltonian of the Tavis-Cummings model in terms of these operators becomes
\begin{align}
H^{(i)}_{NM}=&\alpha\hat{A}_{1}^{\dag}\hat{A}_{1}+\beta\hat{A}_{2}^{\dag}\hat{A}_{2}+\omega_{3}\hat{c}^{\dag}\hat{c} \nonumber\ \\=&\frac{1}{2}\sqrt{\Delta^{2}+4g^{2}\hat{c}^{\dag}\hat{c}}\left(\hat{A}_{1}^{\dag}\hat{A}_{1}-\hat{A}_{2}^{\dag}\hat{A}_{2}\right)+\frac{\Delta}{2}\left(\hat{A}_{1}^{\dag}\hat{A}_{1}+\hat{A}_{2}^{\dag}\hat{A}_{2}\right)+\omega_{3}\hat{c}^{\dag}\hat{c},\label{TCNM}
\end{align}
with $\alpha=\frac{1}{2}\left(\Delta+\Omega\right)$ and $\beta=\frac{1}{2}\left(\Delta-\Omega\right)$.

Since the operators $N_{c}=\hat{c}^{\dag}\hat{c}$, $N_{1}=\hat{A}_{1}^{\dag}\hat{A}_{1}$ and $N_{2}=\hat{A}_{2}^{\dag}\hat{A}_{2}$ commute with each other, we can construct simultaneous eigenstates
\begin{equation}
 |N_{c},N_{1},N_{2}\rangle=|N_{c}\rangle\otimes |N_{1}\rangle\otimes |N_{2}\rangle,
\end{equation}
where the action of the normal-mode operators onto the sates $|N_{1,2}\rangle$ satisfy
\begin{equation}
\hat{A}^{\dag}_{1,2}|N_{1,2}\rangle=\sqrt{N_{1,2}+1}|N_{1,2}+1\rangle, \quad \quad \hat{A}_{1,2}|N_{1,2}\rangle=\sqrt{N_{1,2}}|N_{1,2}-1\rangle.
\end{equation}
Here, the ground state is given by $|N_{c},N_{1},N_{2}\rangle=|0,0,0\rangle$. In this case, the eigenfunctions for the Hamiltonian of the Tavis-Cummings model written as in equation (\ref{TCNM}) are the direct product of the eigenfunctions of the one-dimensional harmonic oscillator. Thus, the corresponding normalized wave function for this case is
\begin{equation}
\psi_{N_{c},N_{1},N_{2}}=\frac{1}{\sqrt{2^{n}N_{c}!N_{1}!N_{2}!}\pi^{3/2}}e^{\rho^{2}/2}H_{N_{c}}(r_{c})H_{N_{1}}(r_{2})H_{N_{2}}(r_{2}),
\end{equation}
where $n=N_{c}+N_{1}+N_{2}$.

Finally, according to these expressions the energy spectrum of the Tavis-Cummings model in terms of the normal-mode operators is given by
\begin{equation}
E^{(i)}_{NM}=\frac{1}{2}\sqrt{(\omega_{2}-\omega_{1})^{2}+4g^{2}N_{c}}\left(N_{1}-N_{2}\right)+\frac{\omega_{2}-\omega_{1}}{2}\left(N_{1}+N_{2}\right)+\omega_{3}N_{c}.\label{ETCNM}
\end{equation}

Therefore, in this Section we solved the Hamiltonian of the Tavis-Cummings model in the interaction picture by using the normal-mode operators, which were used to diagonalize the Hamiltonian.

\subsection{The Tavis-Cummings model and the $SU(2)$ tilting transformation}

Analogous to what was done in Section 3.2, we can rewritte the Hamiltonian of the Tavis-Cummings model as
\begin{equation}
H=\omega_{3}\hat{c}^{\dag}\hat{c}+\frac{(\omega_{1}+\omega_{2})}{2}(\hat{a}^{\dag}\hat{a}+\hat{b}^{\dag}\hat{b})+\frac{(\omega_{2}-\omega_{1})}{2}(\hat{b}^{\dag}\hat{b }-\hat{a}^{\dag}\hat{a})+g(\hat{a}^{\dag}\hat{b}\hat{c}+\hat{a}\hat{b}^{\dag}\hat{c}^{\dag}).
\end{equation}
Now, we introduce the Jordan-Schwinger realization of the $su(2)$ algebra (see Appendix)
\begin{equation}
J_0=\frac{1}{2}\left(a^{\dag}a-b^{\dag}b\right), \quad J_+=a^{\dag}b, \quad J_-=b^{\dag}a,
\end{equation}
in addition to the number operator $N_s=\hat{a}^{\dag}\hat{a}+\hat{b}^{\dag}\hat{b}$. With these operators we obtain the following Hamiltonian
\begin{equation}
H_{su(2)}=\omega_{3}\hat{c}^{\dag}\hat{c}+(\omega_{2}-\omega_{1})\hat{J}_{0}+g(\hat{c}\hat{J}_{+}+\hat{c}^{\dag}\hat{J}_{-})+\frac{(\omega_{1}+\omega_{2})}{2}\hat{N}_{s}.
\end{equation}

Now, in order to remove the ladder operators $J_{\pm}$, we apply the tilting transformation to the eigenvalue equation of this Hamiltonian
\begin{eqnarray}
D^{\dagger}(\xi)\left[\omega_{3}\hat{c}^{\dag}\hat{c}+(\omega_{2}-\omega_{1})\hat{J}_{0}+g(\hat{c}\hat{J}_{+}+\hat{c}^{\dag}\hat{J}_{-})+\frac{(\omega_{1}+\omega_{2})}{2}\hat{N}_{s}\right]D(\xi)D^{\dagger}(\xi)|\Psi\rangle\nonumber\\=E_{su(2)})D^{\dagger}(\xi)|\Psi\rangle,
\end{eqnarray}
where $D(\xi)$ is the $SU(2)$ displacement operator and $\xi=-\frac{1}{2}\theta e^{-i\varphi}$ (see Appendix). If we define the tilted Hamiltonian $H'=D^{\dagger}(\xi)H_{su(2)}D(\xi)$ and the wave function $|\Psi'\rangle=D^{\dagger}(\xi)|\Psi\rangle$, this equation can be written as $\hat{H}'|\Psi'\rangle=E_{su(2)}|\Psi'\rangle$. Moreover, since the operators $N_s$ and $\hat{c}^{\dag}\hat{c}$ commutes whit $\hat{J}_+$ and $\hat{J}_-$, both operators remain unchanged under the tilting transformation. Therefore, we find that the tilted Hamiltonian results to be
\begin{align}
\hat{H}'=&\left[(\omega_{2}-\omega_{1})\left(2\epsilon + 1 \right)-\frac{g\xi^{*}\delta\hat{c}}{|\xi|} -\frac{g\xi\delta\hat{c}^{\dag}}{|\xi|}\right]\hat{J}_{0}\nonumber\\&+\left[\frac{(\omega_{2}-\omega_{1})\delta\xi^{*}}{2|\xi|}+\frac{g\epsilon\xi^{*}\hat{c}}{\xi}+g(\epsilon+1)\hat{c}^{\dag}\right]\hat{J}_{-}\nonumber\\&+\left[\frac{(\omega_{2}-\omega_{1})\delta\xi}{2|\xi|}+g(\epsilon+1)\hat{c}+\frac{g\epsilon\xi\hat{c}^{\dag}}{\xi^{*}}\right]\hat{J}_{+}\nonumber\\&+\frac{\omega_{2}+\omega_{1}}{2}\hat{N}_s+\omega_{3}\hat{c}^{\dag}\hat{c}\label{tilTCsu2}.
\end{align}

By choosing the coherent state parameters $\theta$ and $\varphi$ as
\begin{equation}
\hat{\theta}=\arctan\left(\frac{2g\sqrt{\hat{c}^{\dag}\hat{c}}}{\omega_{2}-\omega_{1}}\right), \quad\quad \hat{\varphi}=i\ln{\left[\frac{2g(2\epsilon+1)\hat{c}}{(\omega_{2}-\omega_{1})\delta} \right]},
\end{equation}
the tilted Hamiltonian $H'$ of equation (\ref{tilTCsu2}) is simplified to
\begin{equation}
\hat{H}'|\Psi'\rangle=\sqrt{(\omega_{2}-\omega_{1})^{2}+4g^{2}\hat{c}^{\dag}\hat{c}}\hat{J}_{0}+\frac{\omega_{2}+\omega_{1}}{2}\hat{N}_s+\omega_{3}\hat{c}^{\dag}\hat{c}|\Psi\rangle=E_{su(2)}|\Psi\rangle.\label{HTC2}
\end{equation}
Since the operators $J_{0}$, $N_{s}$ and $\hat{c}^{\dag}\hat{c}$ commutes with each other, they share common eigenfunctions. Thus, since $N$ is the Hamiltonian of the two-dimensional harmonic oscillator and $\hat{c}^{\dag}\hat{c}$  is the Hamiltonian of the one-dimensional harmonic oscillator, the eigenfunctions of the tilted Hamiltonian $\hat{H}'$ are
\begin{equation}
\Psi'=\langle{z}|n_{a}\rangle\otimes\langle{\rho,\varphi}|n_{l},m_{n}\rangle'=\psi_{c}\otimes\psi'_{\rho,m_{n}},
\end{equation}
where each function is explicitly given by
\begin{equation}
\psi'_{n_{l},m_n}(\rho,\phi)=\frac{1}{\sqrt{\pi}}e^{im_n\phi}(-1)^{n_{l}}\sqrt{\frac{2(n_{l})!}{(n_{l}+m_n)!}}\rho^{m_n}L^{m_n}_{n_{l}}(\rho^{2})e^{-1/2\rho^{2}}\label{Poly5},
\end{equation}
and
\begin{equation}
\psi_{c}(x)=\sqrt{\frac{1}{\pi^{1/4}(2^{n_{c}} n_{c}!)^{1/2}}}e^{-\frac{1}{2}z^2}H_{n_{c}}(z).
\end{equation}
From the action of the operators $\hat{a},\hat{a}^{\dag},\hat{b}$ and $\hat{b}^{\dag}$ on the basis $|n,m_n\rangle$, we have
\begin{equation}
\hat{J}_{0}|n,m_n\rangle=\frac{m_n}{2}|n,m_n\rangle,\nonumber
\end{equation}
\begin{equation}
\hat{N}_s|n,m_n\rangle=(b^{\dag}b+a^{\dag}a)|n,m_n\rangle=(2n_l+m_n)|n,m_n\rangle.
\end{equation}
In this $SU(2)$ representation, the group numbers $j,\mu$ are related with the physical numbers $n_l,m_n$ as $j=n_l+\frac{m_n}{2}$ and $\mu=\frac{m_n}{2}$ \cite{Nos2}. Thus, the energy spectrum of the Tavis-Cummings model described by $su(2)$ Lie algebra is
\begin{equation}
E_{su(2)}=\frac{1}{2}\sqrt{(\omega_{2}-\omega_{1})^{2}+4g^{2}n_{c}}{m_{n}}+(\omega_{2}+\omega_{1})(n_l+\frac{m_n}{2})+\omega_{3}n_{c}.
\end{equation}
It is important to note that this energy spectrum can be related to that obtained with the normal-mode operators (equation (\ref{ETCNM})) by means of a unitary transformation which connects the bases $|N_{c},N_{1},N_{2}\rangle$ and $|n_{a}\rangle\otimes|n_{l},m_{n}\rangle'$.

The eigenfunctions of the Tavis-Cummings model are obtained from the relationship $\Psi=D(\xi)\Psi'$, which only affects the functions $\psi'_{n_{l},m_n}(\rho,\phi)$. Thus, by using equation (\ref{Poly5}) we find that the action of the displacement operator $D(\xi)$ onto $\psi'_{n_{l},m_n}(\rho,\phi)$ is
\begin{align}
\psi_{n_{l},m_{n},\zeta}=&\langle\rho,\varphi|\zeta,n_{l},m_{n}\rangle=\frac{e^{-\frac{\rho^{2}}{2}}}{\sqrt{\pi}}\sum_{s=0}^{n_{l}+n}\frac{\zeta^{s}}{s!}\sum_{n=0}^{n_{l}+m_{n}}\frac{(-\zeta^{*})^{n}}{n!}e^{\frac{\eta}{2}(m_{n}-2n)}e^{i(m_{n}-2n+2s)\varphi}(-1)^{n_{l}+n-s}\nonumber\\&\times\frac{\Gamma(n_{l}+n+1)}{\Gamma(n_{l}+m_{n}-n+1)}\left[\frac{2\Gamma(n_{l}+m_{n}+1)}{\Gamma(n_{l}+1)}\right]^{1/2}\rho^{(m_{n}-2n+2s)}L_{n_{l}+n-s}^{(m_{n}-2n+2s)}(\rho^{2}).\label{F2TC}
\end{align}
Hence, it follows that the eigenfunctions of the Hamiltonian $\hat{H}_{su(2)}$ are given by
\begin{equation}
\Psi=\psi_{c}\otimes\psi_{n_{l},m_{n},\zeta}.
\end{equation}

Therefore, we have studied the Tavis-Cummings model by means of the $SU(2)$ group theory. The solution was obtained by using the tilting transformation and the $SU(2)$ Perelomov number coherent states for the two-dimensional harmonic oscillator.

\section{Expectation value of the Tavis-Cummings Hamiltonian in the $SU(1,1)$ and $SU(2)$ groups.}

In this Section, we are interested in a comparison between the expectation value of the Tavis-Cummings Hamiltonian in terms of the coherent states of the one-dimensional harmonic oscillator, and the coherent states of the $SU(1,1)$ and the $SU(2)$ groups.

Thus, we can introduce the coherent states of the one-dimensional harmonic oscillator $|\alpha\rangle$ and $|\beta\rangle$, and the Perelomov number coherent states of the $SU(1,1)$ group $|\zeta,k,n_{l}\rangle$ and of the $SU(2)$ group $|\zeta,n_{l},m_{n}\rangle$. In terms of these states we can compute the following expectation values for the Tavis-Cummings model
\begin{equation}
\langle H_{su(1,1)}\rangle_{\alpha}=\langle\zeta,k,n_{l}|\otimes\langle\alpha|H_{su(1,1)}|\alpha\rangle\otimes|\zeta,k,n_{l}\rangle,
\end{equation}
\begin{equation}
\langle H_{su(2)}\rangle_{\beta}=\langle\zeta,n_{l},m_{n}|\otimes\langle\beta|H_{su(2)}|\beta\rangle\otimes|\zeta,n_{l},m_{n}\rangle.
\end{equation}
These expressions explicitly are given by
\begin{equation}
\langle H_{su(1,1)}\rangle_{\alpha}=\frac{1}{2}\sqrt{(\omega_{2}+\omega_{3})^{2}-4g^{2}|\alpha|^{2}}\left(n+1\right)+\omega_{1}|\alpha|^{2}+\frac{(\omega_{3}-\omega_{2})}{2}m_{n}-\frac{(\omega_{2}+\omega_{3})}{2},
\end{equation}
with $n=0,1,2,3...$, $m_n=0,\pm1,\pm2,...\pm n $ and $\alpha$ a complex number. Similarly,
\begin{equation}
\langle H_{su(2)}\rangle_{\beta}=\frac{1}{2}\sqrt{(\omega_{2}-\omega_{1})^{2}+4g^{2}|\beta|^{2}}{m_{n}'}+\frac{(\omega_{2}+\omega_{1})}{2}n'+\omega_{3}|\beta|^{2}.
\end{equation}
with $n'=0,1,2,3...$, $m_n'=0,\pm1,\pm2,...\pm n' $ and $\beta$ a complex number.

For the general case, it is more complicated to compare these results and find the values for which both coincide since each one depends on three different field frequencies and the quantum numbers can take different values. Moreover, the group numbers of the $SU(1,1)$ group can take an infinite number of values while the group numbers of the $SU(2)$ are bounded. Therefore, we can work with the isotropic case when $\omega=\omega_{1}=\omega_{2}=\omega_{3}$ and with the assumption $\frac{g^{2}|\alpha|^{2}}{\omega^{2}}\ll1$. With these considerations the expectation values are
\begin{equation}
\langle H_{su(1,1)}\rangle_{\alpha}\approx\left(\omega-\frac{g^{2}|\alpha|^{2}}{\omega}\right)(n+1)+\omega(|\alpha|^{2}-1),
\end{equation}
\begin{equation}
\langle H_{su(2)}\rangle_{\beta}=g|\beta|m_{n}'+\omega n'+\omega|\beta|^{2}.
\end{equation}
For the lowest state of the $SU(2)$ group ($n'=m_{n}'=0$) and the $SU(1,1)$ group ($n=0$), we find that both expectations values matches when
\begin{equation}
|\alpha|=\frac{|\beta|}{\sqrt{1-\left(\frac{g}{\omega}\right)^{2}}},
\end{equation}
where $\alpha$ and $\beta$ are fixed values.
Finally, it is important to note that the mean values of the two descriptions coincide for different values of the quantum numbers ($n$,$m_{n}$), ($n'$,$m_{n}'$) and complex numbers $\alpha$ and $\beta$.

\section{Concluding remarks}

We showed that the Tavis-Cummings model with three modes of oscillation possesses $SU(1,1)$ and $SU(2)$ symmetry and used different algebraic methods to exactly solve this model by using these groups. In the first method, we considered a particular case of the general Tavis-Cummings model and introduced an appropriate Bogoliubov transformation to obtain the energy spectrum. For the second method, we wrote the Tavis-Cummings model in terms of the $su(1,1)$ Jordan-Schwinger realization and used the tilting transformation to diagonalize the general Hamiltonian. With this approach, we were able to obtain the energy spectrum and we showed that the eigenfunctions of this model are the direct product of the one-dimensional harmonic oscillator functions and the $SU(1,1)$ Perelomov number coherent states of the two-dimensional harmonic oscillator. It is important to emphasize that the energy spectrum obtained from the tilting transformation perfectly match with that previously obtained from the Bogoliugov transformation. Moreover, the most general energy spectrum can be properly reduced to that of the one-dimensional, two-dimensional and three dimensional harmonic oscillator with the help of the model parameters.

The $SU(2)$ group theory allowed us to introduce another two algebraic methods, based on the normal-mode operators and the $SU(2)$ tilting transformation. In the normal-mode operators method, we were able to write the Tavis-Cummings model in terms of three number operators. Therefore, the energy spectrum was obtained in a direct way and we showed that the eigenfunctions of the can be expressed as the product of three eigenfunctions of the one-dimensional harmonic oscillator. With the $SU(2)$ tilting transformation we obtained again the energy spectrum of the general model and showed that the eigenfunctions also can be seen as the direct product of the one-dimensional harmonic oscillator functions and the $SU(2)$ Perelomov number coherent states of the two-dimensional harmonic oscillator.

In the last Section, we computed the mean values of the energy of the Tavis-Cummings Hamiltonian by means of the $SU(1,1)$ and $SU(2)$ groups. We showed that these mean values coincide for different values of the quantum numbers of the two-dimensional harmonic oscillator and for different values of the complex numbers of the one-dimensional harmonic oscillator coherent states. Finally, we would like to note that in the $SU(1,1)$ solution, when $\frac{g^{2}}{\omega^{2}}\gg1$ the energy spectrum becomes a complex value. This detail does not happen in the $SU(2)$ solution and the energy spectrum is well defined for any choice of the constants $g$ and $\omega$.

\section*{Acknowledgments}

This work was partially supported by SNI-M\'exico, COFAA-IPN, EDI-IPN, EDD-IPN, SIP-IPN project number $20170632$.

\section{Appendix.}
\subsection{ $SU(1,1)$ Perelomov number coherent states}

Three operators $K_{\pm}, K_0$ close the $su(1,1)$ Lie algebra if they satisfy the commutation relations \cite{Vourdas}
\begin{eqnarray}
[K_{0},K_{\pm}]=\pm K_{\pm},\quad\quad [K_{-},K_{+}]=2K_{0}.\label{com}
\end{eqnarray}
The Casimir operator $K^{2}$ for any irreducible representation of this group is given by
\begin{equation}
K^2=K^2_0-\frac{1}{2}(K_+K_-+K_-K_+)
\end{equation}
The action of these operators on the Fock space states
$\{|k,n\rangle, n=0,1,2,...\}$ is
\begin{equation}
K_{+}|k,n\rangle=\sqrt{(n+1)(2k+n)}|k,n+1\rangle,\label{k+n}
\end{equation}
\begin{equation}
K_{-}|k,n\rangle=\sqrt{n(2k+n-1)}|k,n-1\rangle,\label{k-n}
\end{equation}
\begin{equation}
K_{0}|k,n\rangle=(k+n)|k,n\rangle.\label{k0n}
\end{equation}
\begin{equation}
K^2|k,n\rangle=k(k-1)|k,n\rangle.
\end{equation}
Thus, a representation of $su(1,1)$ algebra is determined by the number $k$, called the Bargmann index. The discrete series
are those for which $k>0$.

The displacement operator $D(\xi)$ is defined in terms of the creation and annihilation operators $K_+, K_-$ as
\begin{equation}
D(\xi)=\exp(\xi K_{+}-\xi^{*}K_{-}),\label{do}
\end{equation}
where $\xi=-\frac{1}{2}\tau e^{-i\varphi}$, $-\infty<\tau<\infty$ and $0\leq\varphi\leq2\pi$.
The so-called normal form of the squeezing operator is given by
\begin{equation}
D(\xi)=\exp(\zeta K_{+})\exp(\eta K_{0})\exp(-\zeta^*K_{-})\label{normal},
\end{equation}
where  $\zeta=-\tanh(\frac{1}{2}\tau)e^{-i\varphi}$ and $\eta=-2\ln \cosh|\xi|=\ln(1-|\zeta|^2)$ \cite{Gerry}.

The $SU(1,1)$ Perelomov coherent states are defined as the action of the displacement operator $D(\xi)$
onto the lowest normalized state $|k,0\rangle$ as \cite{Perellibro}
\begin{equation}
|\zeta\rangle=D(\xi)|k,0\rangle=(1-|\zeta|^2)^k\sum_{n=0}^\infty\sqrt{\frac{\Gamma(n+2k)}{n!\Gamma(2k)}}\zeta^n|k,n\rangle,\label{PCN}
\end{equation}
The $SU(1,1)$ Perelomov number coherent state $|\zeta,k,n\rangle$ is defined as the action of the displacement operator $D(\xi)$ onto an arbitrary
excited state $|k,n\rangle$ \cite{Nos1}
\begin{eqnarray}
|\zeta,k,n\rangle &=&\sum_{s=0}^\infty\frac{\zeta^s}{s!}\sum_{j=0}^n\frac{(-\zeta^*)^j}{j!}e^{\eta(k+n-j)}
\frac{\sqrt{\Gamma(2k+n)\Gamma(2k+n-j+s)}}{\Gamma(2k+n-j)}\nonumber\\
&&\times\frac{\sqrt{\Gamma(n+1)\Gamma(n-j+s+1)}}{\Gamma(n-j+1)}|k,n-j+s\rangle.\label{PNCS}
\end{eqnarray}

The similarity transformations $D^{\dag}(\xi)K_{+}D(\xi)$, $D^{\dag}(\xi)K_{-}D(\xi)$, and
$D^{\dag}(\xi)K_{0}D(\xi)$ of the $su(1,1)$ Lie algebra generators are computed by using the displacement operator $D(\xi)$ and the Baker-Campbell-Hausdorff identity
\begin{equation}
e^{A}Be^{-A}=B+[B,A]+\frac{1}{2!}[[B,A],A]+\frac{1}{3!}[[[B,A]A]A]+...
\end{equation}
These results explicitly are
\begin{equation}
D^{\dag}(\xi)K_{+}D(\xi)=\frac{\xi^{*}}{|\xi|}\alpha K_{0}+\beta\left(K_{+}+\frac{\xi^{*}}{\xi}K_{-}\right)+K_{+},\label{DK+}
\end{equation}
\begin{equation}
D^{\dag}(\xi)K_{-}D(\xi)=\frac{\xi}{|\xi|}\alpha K_{0}+\beta\left(K_{-}+\frac{\xi}{\xi^{*}}K_{+}\right)+K_{-},\label{DK-}
\end{equation}
\begin{equation}
D^{\dag}(\xi)K_{0}D(\xi)=(2\beta+1)K_{0}+\frac{\alpha\xi}{2|\xi|}K_{+}+\frac{\alpha\xi^{*}}{2|\xi|}K_{-},\label{DK0}
\end{equation}
where $\alpha=\sinh(2|\xi|)$ and $\beta=\frac{1}{2}\left[\cosh(2|\xi|)-1\right]$.

A particular realization of the $su(1,1)$ Lie algebra is given by the Jordan-Schwinger operators
\begin{equation}
K_0=\frac{1}{2}\left(a^{\dag}a+b^{\dag}b+1\right), \quad K_+=a^{\dag}b^{\dag},\  \quad K_-= ba,\label{gen11}
\end{equation}
where the two sets of operators $(a,a^{\dag})$ and $(b,b^{\dag})$ satisfy the bosonic algebra
\begin{equation}
[a,a^{\dag}]=[b,b^{\dag}]=1, \quad\quad[a,b^{\dag}]=[a,b]=0.
\end{equation}
If $N_d$ is the difference of the number operators of the two oscillators, then $N_d$ commutes with all
the generators of this algebra and the Casimir operator is given by \cite{vourdasanalytic}
\begin{equation}
K^2=\frac{1}{4}N_d^2-\frac{1}{4}, \quad\quad N_d=b^{\dag}b-a^{\dag}a,\nonumber
\end{equation}
\begin{equation}
[N_d,K_0]=[N_d,K_+]=[N_d,K_-]=0.\label{Nd}
\end{equation}

\subsection{ $SU(2)$ Perelomov number coherent states}

The $su(2)$ Lie algebra is spanned by the generators $J_{+}$, $J_{-}$ and $J_{0}$, which satisfy the commutation relations \cite{Vourdas}
\begin{eqnarray}
[J_{0},J_{\pm}]=\pm J_{\pm},\quad\quad [J_{+},J_{-}]=2J_{0}.\label{com2}
\end{eqnarray}
The Casimir operator $J^2$ in this representation is
\begin{equation}
J^{2}=J_0^2+\frac{1}{2}\left(J_+J_-+J_-J_+\right).
\end{equation}
The action of these operators on the Dicke space states (angular momentum states)\\
$\{|j,\mu\rangle, -j\leq\mu\leq j\}$ is
\begin{equation}
J_{+}|j,\mu \rangle=\sqrt{(j-\mu)(j+\mu+1)}|j,\mu+1 \rangle,\label{j+m}
\end{equation}
\begin{equation}
J_{-}|j,\mu \rangle=\sqrt{(j+\mu)(j-\mu+1)}|j,\mu-1 \rangle,\label{j-m}
\end{equation}
\begin{equation}
J_{0}|j,\mu \rangle=\mu|j,\mu \rangle.\label{j0m}
\end{equation}
\begin{equation}
J^2|j,\mu \rangle=j(j+1)|j,\mu \rangle.
\end{equation}
The displacement operator $D(\xi)$ for this Lie algebra is
\begin{equation}
D(\xi)=\exp(\xi J_{+}-\xi^{*}J_{-}),\label{D}
\end{equation}
where $\xi=-\frac{1}{2}\theta e^{-i\varphi}$. By means of Gaussian decomposition, the normal form of this  operator is given by
\begin{equation}
D(\xi)=\exp(\zeta J_{+})\exp(\eta J_{0})\exp(-\zeta^*J_{-})\label{normal2},
\end{equation}
where  $\zeta=-\tanh(\frac{1}{2}\theta)e^{-i\varphi}$ and $\eta=-2\ln \cosh|\xi|=\ln(1-|\zeta|^2)$.

The $SU(2)$ Perelomov coherent states $|\zeta\rangle=D(\xi)|j,-j\rangle$ are defined as  \cite{Arecchi,Perellibro}
\begin{equation}
|\zeta\rangle=\sum_{\mu=-j}^{j}\sqrt{\frac{(2j)!}{(j+\mu)!(j-\mu)!}}(1+|\zeta|^{2})^{-j}\zeta^{j+\mu}|j,\mu\rangle.\label{PCN2}
\end{equation}
The $SU(2)$ Perelomov number coherent state $|\zeta,j,\mu\rangle$ is defined as the action of the displacement operator $D(\xi)$ onto an arbitrary
excited state $|j,\mu\rangle$
\begin{eqnarray}
|\zeta,j,\mu\rangle &=&\sum_{s=0}^{j-\mu+n}\frac{\zeta^{s}}{s!}\sum_{n=0}^{\mu+j}\frac{(-\zeta^*)^{n}}{n!}e^{\eta(\mu-n)}
\frac{\Gamma(j-\mu+n+1)}{\Gamma(j+\mu-n+1)}\nonumber\\ &&\times\left[\frac{\Gamma(j+\mu+1)\Gamma(j+\mu-n+s+1)}{\Gamma(j-\mu+1)\Gamma(j-\mu+n-s+1)}\right]^{\frac{1}{2}}|j,\mu-n+s\rangle.\label{PNCS2}
\end{eqnarray}

The similarity transformations $D^{\dag}(\xi)J_{+}D(\xi)$, $D^{\dag}(\xi)J_{-}D(\xi)$, and $D^{\dag}(\xi)J_{0}D(\xi)$ of the $su(2)$ Lie algebra generators are computed by using the displacement operator $D(\xi)$ and the Baker-Campbell-Hausdorff identity
\begin{equation}
D^{\dag}(\xi)J_{+}D(\xi)=-\frac{\xi^{*}}{|\xi|}\delta J_{0}+\epsilon\left(J_{+}+\frac{\xi^{*}}{\xi}J_{-}\right)+J_{+},
\end{equation}
\begin{equation}
D^{\dag}(\xi)J_{-}D(\xi)=-\frac{\xi}{|\xi|}\delta J_{0}+\epsilon\left(J_{-}+\frac{\xi}{\xi^{*}}J_{+}\right)+J_{-},
\end{equation}
\begin{equation}
D^{\dag}(\xi)J_{0}D(\xi)=(2\epsilon+1)J_{0}+\frac{\delta\xi}{2|\xi|}J_{+}+\frac{\delta\xi^{*}}{2|\xi|}J_{-},
\end{equation}
where $\delta=\sin(2|\xi|)$ and $\epsilon=\frac{1}{2}\left[\cos(2|\xi|)-1\right]$.

The Jordan-Schwinger realization of the $su(2)$ algebra is
\begin{equation}
J_0=\frac{1}{2}\left(a^{\dag}a-b^{\dag}b\right), \quad J_+=a^{\dag}b, \quad J_-=b^{\dag}a,\label{gen}
\end{equation}
where again the two sets of operators $(a, a^{\dag})$ and $(b, b^{\dag})$ satisfy the bosonic algebra.
It is important to note that, besides the Casimir operator, there is another operator $N_s$ (called the number operator)
which commutes with all the generators of the $su(2)$ algebra.

The Casimir operator $J^2$ for this realization is written in terms of this operator $N$ and is given by
\begin{equation}
J^2=\frac{N_s}{2}\left(\frac{N_s}{2}+1\right),\quad\quad N_s=a^{\dag}a+b^{\dag}b,\nonumber
\end{equation}
\begin{equation}
[N_s,J_+]=[N_s,J_-]=[N_s,J_z]=0.
\end{equation}

\end{document}